\documentclass[fleqn,twoside]{article}
\usepackage{espcrc2}
\usepackage{graphicx}
\usepackage[figuresright]{rotating}

\newcommand{\AmS}{{\protect\the\textfont2 A\kern-.1667em\lower.5ex\hbox{M}\kern-.125emS}}
\newcommand{\nue}{\nu_{e}}
\newcommand{\num}{\nu_{\mu}}
\newcommand{\nut}{\nu_{\tau}}
\newcommand{\nus}{\nu_{sterile}}

\title{Atmospheric neutrino oscillations with MACRO}

\author{M. Sioli\address[MCSD]{Physics Department of the University and INFN,
Viale C. Berti Pichat 6/2, I-40127, Bologna, Italy},
for the MACRO Collaboration\thanks{List of authors and institutions: see Ref \cite{macro}}}
       
\begin{document}

%

\begin{abstract}
\vspace{-3.5cm}
\hspace{12cm}
\bf{DFUB 12/2002}

\hspace{12cm}
\bf{Bologna, 04/11/02}
\vspace{2.2cm}
\begin{center}
  {\bf Contribution to the XII ISVHECRI, CERN, Geneva, 15-20 July 2002}
\end{center}
\normalfont
We present the latest results on the study of atmospheric neutrino oscillations 
with the MACRO detector at Gran Sasso. Two sub-samples of events have been analysed,
both in terms of absolute flux and zenith angle distribution:
high energy events (with $\langle E_{\nu} \rangle \simeq$ 50 GeV) and
low energy events (with $\langle E_{\nu} \rangle \simeq$ 4 GeV). 
The high energy sample has been used also to check the $\num \leftrightarrow \nu_{sterile}$ 
oscillation hypothesis and to estimate neutrino energies using Multiple Coulomb Scattering (MCS) informations.
All these analyses are mutually consistent and strongly favour the $\num \leftrightarrow \nut$ 
oscillation hypothesis with maximal mixing and $\Delta m^{2} = 2.5 \cdot 10^{-3}$eV$^{2}$.

\vspace{1pc}
\end{abstract}

\maketitle

\section{INTRODUCTION}
\begin{figure}[t!]
\includegraphics*[width=17pc]{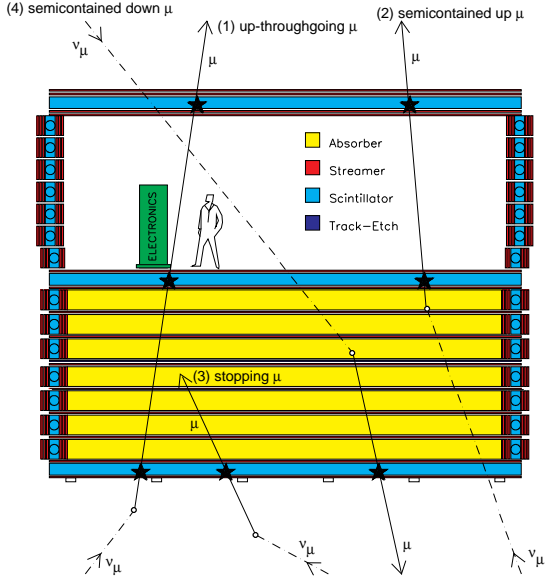}
\vspace{-0.7cm}
\caption{Schematic picture of the different topologies of $\nu$-induced events in MACRO.
From the left: 1) upward throughgoing muons, 2) internal upgoing (IU) muons, 
3) upgoing stopping muons (UGS) and 4) internal downgoing (ID) muons. 
Samples (3+4) are studied together, see text.}
\label{f:topo}
\end{figure}

The MACRO detector \cite{macro} was located in the Gran Sasso Laboratory (Italy) and allowed the study
of atmospheric neutrinos detecting upgoing muons produced in CC interactions inside or 
around the detector. It used scintillation counters for tagging events by 
time-of-flight (TOF) measurements, and limited streamer tubes for tracking.
Three categories of $\num$-induced muon events were studied (see Fig. \ref{f:topo}): 
1) upthroughgoing muons, produced in the rock below the detector and crossing 
the whole apparatus, with an average energy $\langle E_{\nu} \rangle \simeq$ 50 GeV; 
2) internal upgoing (IU) muons, produced inside and leaving the detector from above;
3) upgoing stopping muons (UGS), produced below and stopping inside the detector plus
internal downgoing (ID) muons, produced inside and leaving the detector from below.
Since at least two scintillation counters are needed to perform a TOF measurement, the events of 
samples UGS and ID are studied on a topological basis and are indistinguishable. 
Therefore, they are studied together.
The average neutrino energy for samples 2 and 3 is about $\simeq$ 4 GeV, and we collectively 
refer to these events as to the Low Energy (LE) sample \cite{le}, while events from sample 1 belong to the 
High Energy (HE) sample \cite{he}.

\section{HIGH ENERGY SAMPLE}
Events in the HE sample have been selected by measuring the TOF between two layers of scintillators.
For events crossing three layers of scintillators, a linear fit of the times as a function of 
the path lengths has been performed, in order to reduce possible fake events. A detailed tuning of
the TOF measurements provided a reduction factor of $\sim 10^{7}$ of the downgoing atmospheric
muons. Other sources of background have been reduced by applying a cut on the matching positions 
between scintillators and streamer tubes. 
Soft upgoing hadrons coming from photonuclear interactions of muons outside the detector
could mimic an upgoing muon: to reduce this background we require that each upgoing muon crosses
at least 200 g/cm$^{2}$ of absorber in the lower part of the detector \cite{spurio}. 
Muons coming from particular
azimuthal regions, where the rock amount is not sufficient to reduce the number of downgoing muons, 
have been discarded. 863 events survived these cuts (809 after background subtraction), corresponding
to a livetime of 6.16 yrs.

Monte Carlo (MC) simulation of upgoing muon events has been performed using the $\nu$ flux computed
by the Bartol group \cite{bartol}, neutrino cross sections from Ref. \cite{cs} and muon energy loss
in the rock from calculations of Ref. \cite{eneloss}. The overall theoretical uncertainty
on the absolute upgoing muon flux is 17\%, while the systematic error on the shape of the angular
distribution is 5\%.

\begin{figure}[t!]
\includegraphics*[width=17pc]{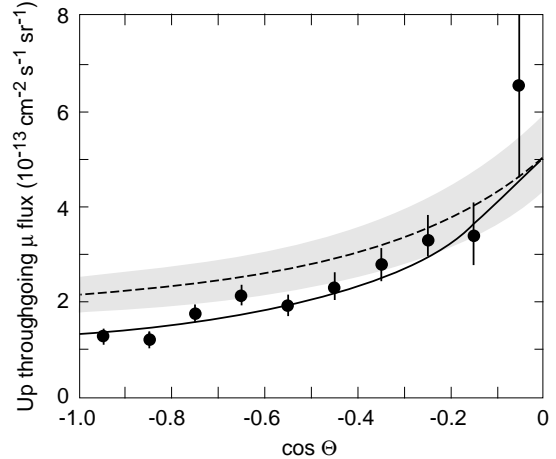}
\vspace{-0.7cm}
\caption{Angular distribution of upthroughgoing muons. Black points are data, with statistical
and systematic errors summed in quadrature. Shaded region is Monte Carlo prediction, assuming 
no oscillations and with a 17\% error on the scale. The solid line is the Monte Carlo prediction
assuming $\num \leftrightarrow \nut$ oscillations with maximal mixing and 
$\Delta m^{2} = 2.5 \cdot 10^{-3}$eV$^{2}$.}
\label{f:passanti}
\end{figure}

Compared to MC predictions the number of measured events is small and the shape of the angular
distribution is different, Fig. \ref{f:passanti}.
The ratio between the observed number of events and the MC prediction is 
$R_{HE} = 0.721 \pm 0.026_{stat} \pm 0.043_{sys} \pm 0.123_{th}$.
The shapes of the experimental distribution and MC predictions assuming no oscillations, analysed in 
terms of $\chi^2$, give an agreement probability of 0.2\%. 
Assuming $\num \leftrightarrow \nut$ oscillations, the angular distribution and the absolute number of events
give a best fit for maximal mixing and $\Delta m^{2} = 2.5 \cdot 10^{-3}$eV$^{2}$, 
with a $\chi^2$ probability of 66\%. 
The result of the fit is shown as a solid line in Fig. \ref{f:passanti}.

\subsection{$\num \leftrightarrow \nu_{sterile}$ oscillations}
\begin{figure}[t!]
\includegraphics*[width=17pc]{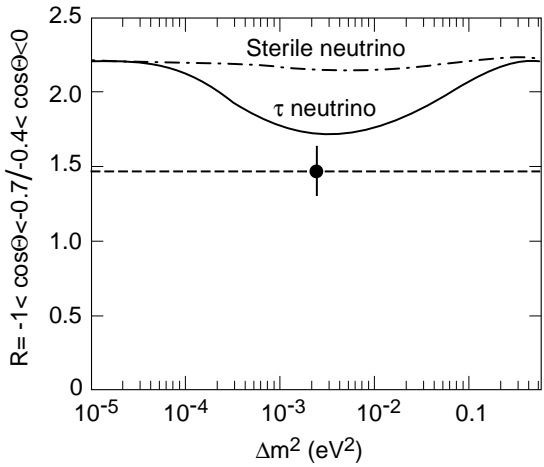}
\vspace{-0.7cm}
\caption{ratio of the number of events with -1 $< \theta <$ -0.7 with respect to the number of events with 
-0.4 $< \theta <$ 0. The black point is the measured value (with error bar), the solid lines are the
prediction for $\num \leftrightarrow \nut$ and $\num \leftrightarrow \nus$ cases.}
\label{f:sterile}
\end{figure}

The weak potential of $\nue$ and $\nus$ with matter is different from the $\num$ and $\nut$ potentials.
On the other hand, $\num$ and $\nut$ potentials are equal: this difference translates in a distortion of the oscillation
pattern (and hence of the angular distribution) in the $\num \leftrightarrow \nus$ oscillation with respect
to the $\num \leftrightarrow \nut$ case. Matter effects become important when 
$E_{\nu} / |\Delta m^2| \geq 10^{-3}$ GeV/eV$^{2}$, i.e. for HE events.
A detailed analysis \cite{sterile} has shown that the best estimator to disentangle the two hypotheses is the
ratio $R$ between the number of events with -1 $< \Theta <$ -0.7 and the number of events with 
-0.4 $< \Theta <$ 0, where $\Theta$ is the zenith angle.
In this ratio, most of the theoretical uncertainties cancel and the overall uncertainty,
combined with the experimental systematic error, is $\sim$ 7\%.
The measured ratio is $R = 1.48 \pm 0.13_{stat} \pm 0.10_{sys}$. This result is shown in Fig. \ref{f:sterile}
together with the MC prediction, as a function of the $\Delta m^2$. In the best fit point, the expected
values for $\num \leftrightarrow \nut$ and $\num \leftrightarrow \nus$ are $R_{\tau} = 1.72$
and $R_{\tau} = 2.16$ respectively. The ratio of the probabilities to obtain values as low as the observed
one is $R_{prob} = 157$, therefore $\num \leftrightarrow \nus$ can be excluded with respect $\num \leftrightarrow \nus$
at 99\% CL.

\subsection{Energy estimate by means MCS}
The limited streamer tube system of the MACRO detector has been used in order to estimate muon energies
(and hence neutrino energies) using MCS informations. A first analysis, which used
streamer tubes in digital mode (with a spatial resolution of $\sigma \simeq 1$ cm),
successfully showed the feasibility of the method \cite{bakari}. In order to improve 
the sensitivity of the analysis to higher neutrino energies, we used the streamer tubes in
drift mode, improving the resolution by a factor $\sim$ 3.5 \cite{tecnico}. Two dedicated
test beams at CERN PS/SPS checked electronics, detector performances and analysis tools. 
Starting from MCS sensitive variables, a MC trained neural network was used for 
muon (and neutrino) energy reconstruction.
These results allowed to increase the sensitivity of the analysis up to neutrino energies of $\sim$ 100 GeV.
We separated the HE sample in four different energy regions: the results show that the agreement between data 
and MC predictions follows the energy dependence expected in the oscillation hypothesis with the
parameters given in the previous section \cite{scapeio}.

\begin{figure}[t!]
\includegraphics*[width=17pc]{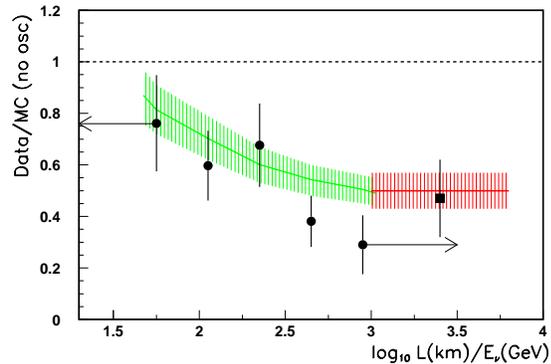}
\vspace{-0.7cm}
\caption{Ratio $DATA/MC(no-osc)$ as a function of the estimated $log_{10} L/E_{\nu}$ (black points).
The solid line is the ratio $MC (sin^{2} 2\theta =1, \Delta m^{2} = 2.5 \cdot 10^{-3}$eV$^{2}$)/$MC(no-osc)$,
with corresponding systematic error (see text). The black square refers to IU events.}
\label{f:lsue}
\end{figure}

The distance travelled by neutrinos inside the Earth (reconstructed with a precision of $\sim$ 3\%)
was used to estimate the $L/E_{\nu}$ on a event-by-event basis. This is shown in Fig. \ref{f:lsue},
where the ratio $DATA/MC(no osc)$ as a function of the estimated log$_{10} L/E_{\nu}$ is plotted.
Errors include uncertainties both on the flux and on the shape. MC prediction is also reported,
with the same parameters of the standard analysis, showing that the trend of the data is the one we 
expect in the $\num \leftrightarrow \nut$ oscillation hypothesis.

\section{LOW ENERGY SAMPLE}
\begin{figure}[t!]
\includegraphics*[width=17pc]{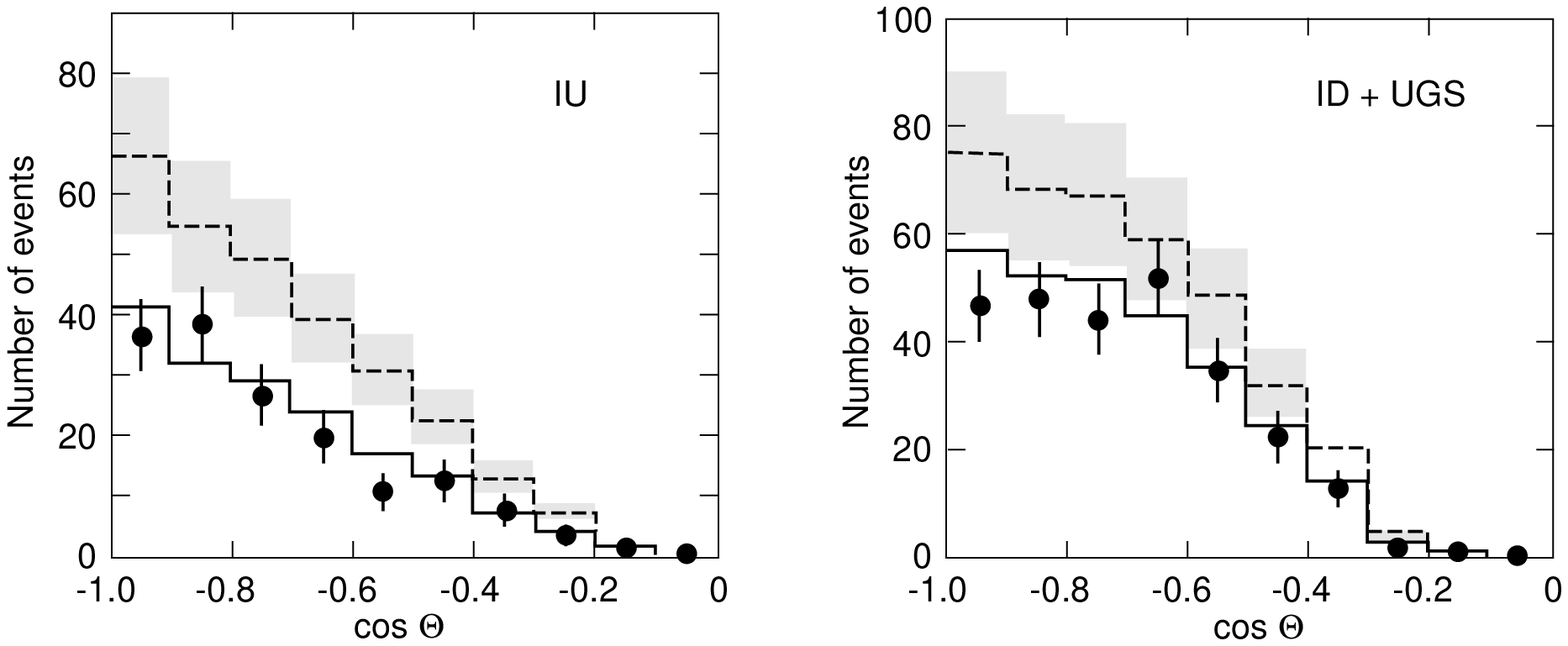}
\vspace{-0.7cm}
\caption{Angular distribution of IU events (a) and ID+UGD events (b). Black points are data, with statistical
and systematic errors summed in quadrature. Dashed lines are Monte Carlo predictions, assuming 
no oscillations and with a 17\% error on the scale (shaded region). Solid lines are the Monte Carlo predictions
assuming $\num \leftrightarrow \nut$ oscillations with maximal mixing and $\Delta m^{2} = 2.5 \cdot 10^{-3}$eV$^{2}$.}
\label{f:semicont}
\end{figure}

IU and ID+UGS events are produced in CC interactions inside the detector, with a small contamination 
of NC and $\nue$ events ($\sim$ 13\% and $\sim$ 10\% respectively). IU events have been selected by
TOF measurement and by topological criteria, e.g. the requirement of the interaction vertex in the
fiducial volume of the detector. The ID+UGS sample has been selected only with topological criteria.
Since no timing informations were available for these events, particular attention has been devoted
in order to reject events not perfectly confined in the detector. These goals have been reached also
by means of visual scanning with the Event Display.
After these selections, we remained with 154 and 262 events for the IU and ID+UGS, corresponding
to 5.8 yrs and 5.6 yrs of livetime respectively. In Fig. \ref{f:semicont} we show the angular
distributions of these two samples and the MC predictions with and without the oscillation hypotheses.
A conservative value (25\%) of the theoretical uncertainty on the absolute scale is reported.
There is a good agreement between data and MC prediction with the oscillation parameters given by
the HE sample analysis.

\section{CONCLUSIONS}
\begin{figure}[t!]
\includegraphics*[width=17pc]{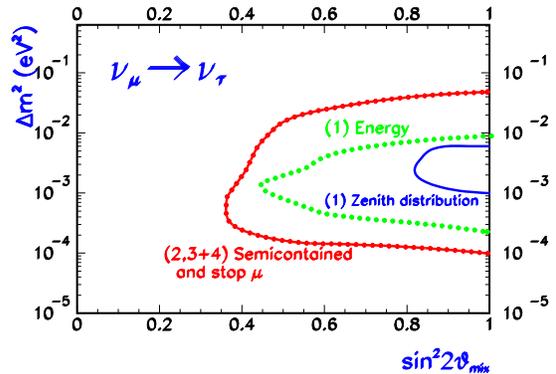}
\vspace{-0.7cm}
\caption{90\% CL contours for allowed region in the ($sin^{2} 2\theta, \Delta m^{2}$) plane.
Curves 1 and 2 refer to upthroughgoing muons using flux+angular distribution and energy
estimate respectively. The curve 3 refers to the LE sample.}
\label{f:contour}
\end{figure}

We presented different and independent analyses performed on muon events induced by atmospheric neutrinos.
All the analyses are mutually consistent and strongly favour the $\num \leftrightarrow \nut$
oscillation hypothesis with maximal mixing and $\Delta m^{2} = 2.5 \cdot 10^{-3}$eV$^{2}$.
Absolute flux measurements, shape distributions and energy estimate by MCS were used in order
to compute allowed regions in the space of oscillation parameters. This is shown in Fig. \ref{f:contour},
where the corresponding contours are computed according to a $\chi^2$ analysis using the prescriptions
of Ref. \cite{fc}. A global analysis of all the samples is in progress.

\end{document}